\pdfoutput=1%
\documentclass[a4paper,american,floatfix,pdftex,superscriptaddress,twoside,%
aip,rsi,%
citeautoscript,
reprint,twocolumn,
]{revtex4-2}%

\PassOptionsToPackage{unicode}{hyperref} 
\usepackage{amsfonts,amsmath,amssymb}
\usepackage{graphicx}
\usepackage[T1]{fontenc}
\usepackage[utf8]{inputenc}
\usepackage{hypernat}
\usepackage[ams,todos]{CMImacros}
\usepackage{verbatim}
\usepackage{chemformula}
\usepackage{mhchem}

\graphicspath{{./fig/}{./}} 
\newcommand{\cfeldesy}{\affiliation{Center for Free-Electron Laser Science CFEL, Deutsches
      Elektronen-Synchrotron DESY, Notkestr. 85, 22607 Hamburg, Germany}}
\newcommand{\uhhcui}{\affiliation{Center for Ultrafast Imaging, Universität Hamburg, Luruper
      Chaussee 149, 22761 Hamburg, Germany}}%
\newcommand{\uhhphys}{\affiliation{Department of Physics, Universität Hamburg, Luruper Chaussee 149,
      22761 Hamburg, Germany}}
\newcommand{\jlu}{\altaffiliation[Current address:~]{Institute of Atomic and Molecular Physics, Jilin
      University, Changchun 130012, China}}%

\newcommand{\jkemail}{\email[Email:~]{jochen.kuepper@cfel.de}}%
\newcommand{\stemail}{\email[Email:~]{sebastian.trippel@cfel.de}}%
\newcommand{\cmiweb}{\homepage[\\Website:~]{https://www.controlled-molecule-imaging.org}}%

\begin{document}
\title{A versatile and transportable endstation for controlled molecule experiments}

\author{Wuwei~Jin}\cfeldesy\uhhphys%
\author{Hubertus~Bromberger}\cfeldesy%
\author{Lanhai~He}\jlu\cfeldesy%
\author{Melby~Johny}\cfeldesy\uhhphys\uhhcui%
\author{Ivo~S.~Vinkl\'{a}rek}\cfeldesy
\author{Karol~Długołęcki}\cfeldesy
\author{Andrey~Samartsev}\cfeldesy
\author{Francesca Calegari}\cfeldesy\uhhphys\uhhcui
\author{Sebastian~Trippel}\stemail\cfeldesy\uhhcui%
\author{Jochen~Küpper}\jkemail\cmiweb\cfeldesy\uhhphys\uhhcui

\begin{abstract}\noindent%
   We report on a new versatile transportable endstation for controlled molecule (eCOMO) experiments
   providing a combination of molecular beam purification by electrostatic deflection and
   simultaneous ion and electron detection using velocity-map imaging (VMI). The $b$-type
   electrostatic deflector provides spatial dispersion of species based on their
   effective-dipole-moment-to-mass ratio. This enables selective investigation of molecular
   rotational quantum states, conformers, and molecular clusters. Furthermore, the double-sided VMI
   spectrometer equipped with two high-temporal-resolution event-driven Timepix3 cameras provides
   detection of all generated ions independently of their mass-over-charge ratio and electrons. To
   demonstrate the potential of this novel apparatus, we present experimental results from our
   investigation of carbonyl sulfide (OCS) after ionization. Specifically, we provide the
   characterization of the molecular beam, electrostatic deflector, and electron- and ion-VMI
   spectrometer. The eCOMO endstation delivers a platform for ultrafast dynamics studies using a
   wide range of light sources from table-top lasers to free-electron-laser and
   synchrotron-radiation facilities. This makes it suitable for research activities spanning from
   atomic, molecular, and cluster physics, over energy science and chemistry, to structural biology.
\end{abstract}
\date{\today}%
\maketitle%

\section{Introduction}
\label{sec:introduction}
The advancement of synchrotron and x-ray free-electron laser (XFEL) radiation sources, such as those
at DESY (FLASH), the linac coherent light source (LCLS), the free-electron laser radiation for
multidisciplinary investigations (FERMI), the SPring-8 Angstrom compact free-electron laser (SACLA),
the x-ray free-electron laser at the Paul Scherrer Institute (SwissFEL), the European x-ray
free-electron laser (EuXFEL), and the Shanghai soft x-ray free-electron laser (SXFEL), substantially
pushes the boundaries of existing research areas and breaks the ground for completely new
fields~\cite{Neutze:Nature406:752, Feldhaus:JPhysB9:S799, Feldhaus:JPB46:164002,
   Altarelli:XFEL-TDR:2006, Neutze:COISB22:651, Chapman:NatPhys2:839, Erk:Science345:288,
   Kuepper:PRL112:083002, Chapman:PTRSB369:20130313, Rudenko:Nature546:129, Sobolev:CommPhys3:97,
   ONail:PRL125:073203, Li:Science375:285, Boll:NatPhys18:423}. These sources generate ultrashort
x-ray pulses spanning from tens of femtoseconds down to the attosecond
range~\cite{Pellegrini:RMP88:015006, Huang:Innovation2:100097, Hartmann:NatPhoton12:215}.
Furthermore, they provide a high peak brilliance, coherence and photon energies ranging from 5 to
250~keV~\cite{Decking:NatPhoton14:391, Ackermann:NatPhoton1:336, Altarelli:XFEL-TDR:2006,
   Tschentscher:ApplSci, Schoenlein:SLACR1053:2015, Young:JPB51:032003, Tiedtke:NJP11:023029,
   Emma:NatPhoton4:641, Franz:SynchRadNews19:25}. Table-top laser systems offer unique advantages in
other areas: These sources include variable-wavelength optical-parametric-amplifier (OPA)-based
systems~\cite{Cerullo:RSI:74:1, Musheghyan:JPB:185402}, high power lasers with repetition rates up
to 1~MHz using optical parametric chirped-pulse amplification (OPCPA)~\cite{Mecseki:OptLett44:1257,
   Wolter:PRX5:021034}, and lasers with attosecond pulse durations~\cite{Krausz:RMP:81:163,
   Nisoli:CR117:10760}. Each photon source has unique properties, functions, and parameters that
make it suitable for different investigations, ranging from ultrafast studies of chemical reactions
over bio-molecular dynamics to materials and life
science~\cite{Calegari:Science346:336,Strueder:NIMA614:483, Chapman:NatMater8:299,
   Chapman:Nature470:73, Woodhouse:NatComm11:741, Barty:ARPC64:415}. This progress led to particular
needs on user-side applications, notably atomic-level-resolution molecular dynamics experiments
based on ultra-cold molecular beam methods~\cite{Young:Nature466:56, Rudek:NatPhoton6:858,
   Feldhaus:JPB46:164002, Eichmann:Science369:1630, Li:PRL127:093202, Fehre:PRL127:103201,
   Lee:NatComm12:6107, Kuepper:PRL112:083002}.

Molecular beams are of significant importance in physical chemistry and molecular physics as they
provide a unique opportunity to obtain fundamental insights into molecular photophysics and chemical
processes~\cite{Dunoyer:ComptRend152:592, Fraser:MolRay, Herschbach:DFS33:149, Carr:NJP11:055049,
   Meerakker:NatPhys4:595}. Whereas substantial research was conducted on isolated small molecules
and atoms, investigating more complex molecular systems with molecular-level insight is still
restricted by many experimental challenges~\cite{Farnik:JPCL14:287, Boll:NatPhys18:423,
   Chang:IRPC34:557}. One of these challenges is to generate well-defined ensembles of the reactant
particles. In this regard, many molecular beams usually contain mixtures of isolated molecules,
various molecular clusters, or multiple isomers, \eg, of chiral molecules or large biomolecules. The
ability to select a particular species plays a crucial role in gaining a deeper understanding of its
importance in chemical behavior that also impacts a wide range of biological
processes~\cite{Milner:PRL122:223201, Beaulieu:Science358:1288, Onvlee:NatComm13:7462,
   Johny:PCCP26:13118}. To overcome this obstacle, the electrostatic deflector was developed to a
achieve spatial separation of different neutral species in ultra-cold ($T_{\text{rot}}<1$~K)
molecular beams, ultimately obtaining pure particle ensembles~\cite{Filsinger:PRL100:133003,
   Filsinger:ACIE48:6900, Trippel:PRA86:033202, Chang:IRPC34:557, Horke:ACIE53:11965,
   Teschmit:ACIE57:13775}.

A second challenge is recording the molecular dynamics using experimentally accessible observables.
The reaction microscope (REMI) technique, for example, provides complete, correlated, 3D velocity
information of all particles detected in coincidence~\cite{Ullrich:RPP66:1463, Wolter:PRX5:021034,
  Boll:NatPhys18:423}. However, it is limited by low count-per-shot rates, which can lead to
poor statistics. High-count-rate approaches such as VMI~\cite{Eppink:RSI68:3477}, provide a
significant improvement in that respect. A significant drawback of VMI in high count rate conditions
however is that one cannot rely on coincidence detection to extract, \eg, information about
different electrons and ions coming from the same target molecule, although covariance mapping
methods~\cite{Frasinski:Science246:1989} can partially overcome this problem. Lastly, for VMIs,
photoelectron photoion coincidence (PEPICO) spectroscopy that is synchronizing VMI- and
time-of-flight (TOF) information, is also challenging~\cite{Bodi:RSI83:083105}. These issues so far
limit access to the full kinematics of single reactions and, therefore, systematic studies of the
underlying physical mechanisms.

This article provides a comprehensive overview of the newly established \underline{e}ndstation for
\underline{co}ntrolled \underline{mo}lecule experiments (eCOMO). It addresses the aforementioned
challenges of the current experimental research. eCOMO provides pure, cold, dense, and pulsed
molecular beams of selected species using Even-Lavie valves~\cite{Even:EPJTI2:17} in combination
with the electrostatic deflector~\cite{Chang:IRPC34:557}. The endstation is equipped with a
double-sided VMI spectrometer combined with high spatio-temporal-resolution Timepix3 cameras
combined with the software package PymePix~\cite{AlRefaie:JINST14:P10003, Bromberger:JPB55:144001}.
This event-driven approach enables us overcome the limitations of standard, frame-based, camera
technology. It allows for quickly and accurately reconstructing of the full 3D-momentum distribution
of all created ions for advanced imaging experiments with high event rates~\cite{Zhao:RSI88:113104,
   Bromberger:JPB55:144001, Tremsin:RadMea130:106228, Wood:JASMS33:2328}. Moreover, this also meets
the high-frequency data acquisition requirements of various light sources with high repetition rates
mentioned earlier~\cite{Decking:NatPhoton14:391, Bromberger:JPB55:144001}.

To demonstrate the capabilities of the novel system, we present the results of its commissioning
with an optical laser in our laboratory. We used the fragmentation of OCS after strong-field
ionization as a benchmark to characterize the molecular beam, the electrostatic deflector, and the
electron-and-ion-VMI spectrometer.

\section{Apparatus description}
\label{sec: Aparatus description}
\subsection{Vacuum system}
\label{sec:Vacuum system}
\begin{figure*}
   \includegraphics[width=\linewidth]{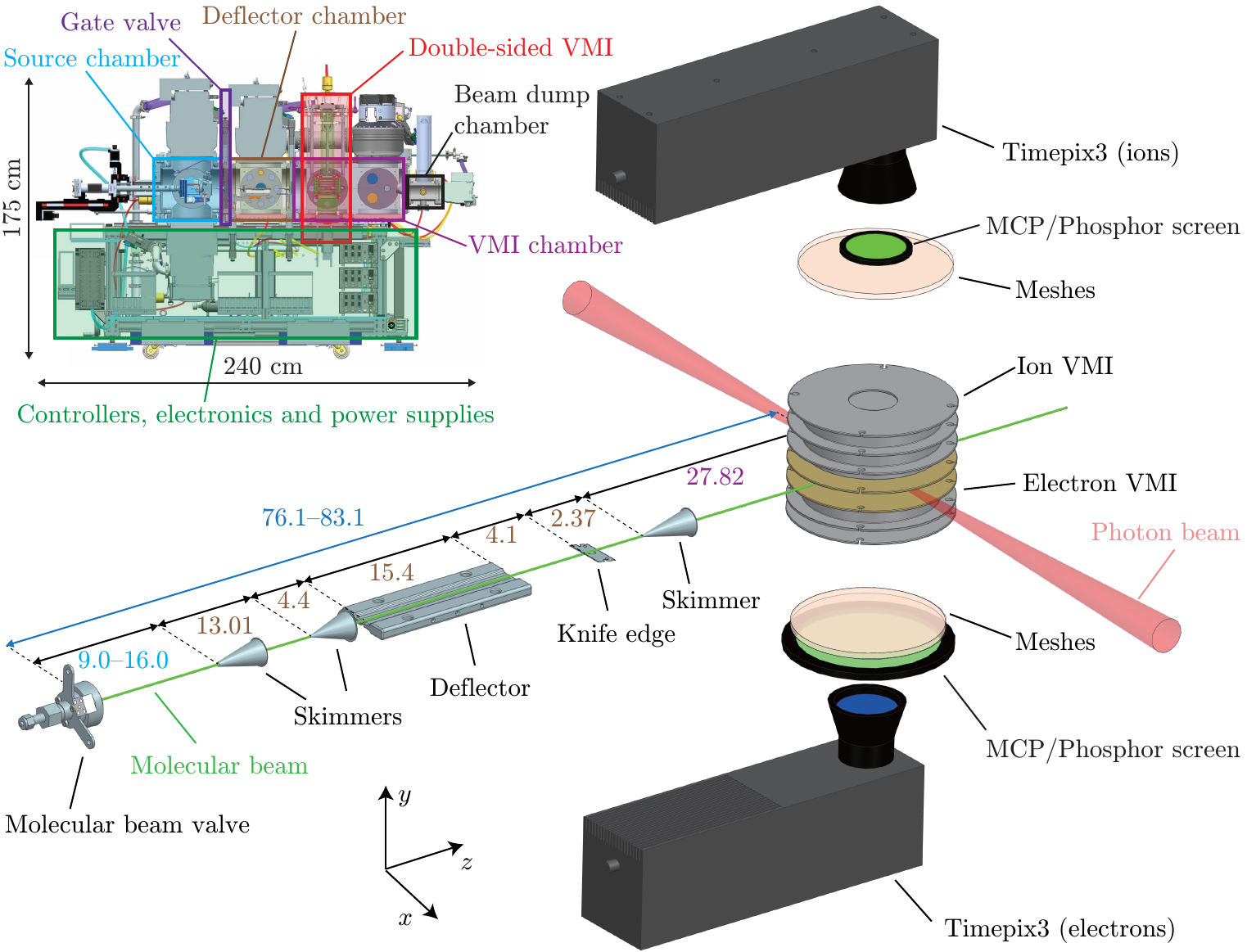}
   \caption{Sketch of the experimental setup consisting of a molecular beam valve, an electrostatic
      $b$-type deflector, a double-sided VMI, and two Timepix3 cameras. The inset provides a
      side view from a CAD model, where colored frames depict the different parts of the setup. All
      distances are provided in cm. See text for details.}
   \label{fig:experimental_setup}
\end{figure*}

The vacuum setup, shown schematically in the upper left corner of \autoref{fig:experimental_setup},
consists of four differentially pumped vacuum chambers housing the molecular beam source, the
electrostatic $b$-type deflector~\cite{Kienitz:JCP147:024304}, the double-sided VMI spectrometer,
and the molecular beam dump.
All chambers are mounted on rails to allow easy translation of the setup along the $z$-axis, which
is parallel to the molecular beam propagation direction. The source chamber is pumped with two
turbomolecular pumps (Pfeiffer Vacuum ATH2303~M), which results in a total pumping speed of
$\ordsim3600$~l/s for helium. The source chamber is equipped with two quick-access DN250~CF doors
(Pfeiffer Vacuum 420KTU250), enabling easy sample replacement in the Even-Lavie valve sample
container~\cite{Even:EPJTI2:17}. Furthermore, the source chamber is separated from the deflector
chamber with home-built automatic DN250~CF gate valve~\cite{Kuepper:RSI77:016106} to allow sample
changes without breaking the vacuum of the deflector chamber and the spectrometer. The deflector
chamber is also pumped with a turbomolecular pump with a pumping speed on the order of
$\ordsim1800$~l/s for helium (Pfeiffer Vacuum ATH2303~M). Next, the VMI chamber is pumped by two
turbomolecular pumps (Pfeiffer Vacuum PM~P06~302 and PM~P06~312) resulting in an overall pumping
speed of $\ordsim4000$~l/s for helium. Finally, the beam dump is pumped with a turbomolecular pump
(Pfeiffer Vacuum PM~P04~024) resulting in a pumping speed of $\ordsim470$~l/s for helium. In
addition, a 250~mm diameter copper shield connected to a dewar for liquid nitrogen cooling is
located in the VMI chamber to increase the pumping speed for molecules with freezing points above
the liquid nitrogen temperature. The Dewar is designed such that liquid nitrogen must be refilled
every $\ordsim7$ hours to keep the shield cold.

In order to achieve better vacuum, the VMI chamber can be baked. All non-metallic components of the
VMI spectrometer are made of glass ceramic (Macor) with an operating temperature of up to
\celsius{\larger600}. The temperature limiting components of the setup are the O-ring and copper
seals with a maximum temperature of \celsius{200} each. It must also be noted that the pumps must
not exceed a temperature of \celsius{120} with possible additional air cooling. In addition, the
pressure gauges must not be operated above \celsius{55}. So far, baking was performed at
\celsius{120}. After one week of baking and subsequent cooldown, a pressure of
$3\times10^{-11}$~mbar was achieved.

The first pre-vacuum line that provides vacuum to the source chamber and to the second low pressure
pre-vacuum line, is pumped by a pre-vacuum pump (Pfeiffer Vacuum PR P00 009) with a nominal pumping
speed of $\ordsim 130~\mathrm{m^3/h}$. The second pre-vacuum line is pumped by a turbomolecular pump
(Pfeiffer Vacuum PM~P03~900) with a pumping speed of 255~l/s for helium. The second pre-vacuum line
provides vacuum to the deflector-, VMI-, and beam dump chambers. This line is needed to compensate
for the low helium compression ratio given for, \eg, the pumps of the VMI chamber by
$3\times10^{5}$. When operating the molecular beam valve at a repetition rate of 100~Hz, the typical
pressures in the source-, deflector-, VMI-, and beam dump chambers are $\ordsim 10^{-6}$, $\ordsim
10^{-8}$, $\ordsim 10^{-10}$, and $\ordsim 10^{-10}$~mbar, respectively. Under such conditions, the
typical pressure in the first and second pre-vacuum lines is $\ordsim 10^{-3}$ and $\ordsim
10^{-7}$~mbar, respectively.

Further information on the vacuum system including the layout of gas
lines, the interlock system with its logic to control the gate valves, and voltage power supplies,
is provided in the supplementary information. The primary purpose of the interlock system is to
ensure operation monitoring of the eCOMO vacuum system and to automatically protect the setup and
spectrometer in case of sudden emergency situations.

\subsection{Molecular beam setup}
\label{sec: Molecular beam setup}

The main part of \autoref{fig:experimental_setup} shows the molecular beam setup. All molecular beam
components including the valve, skimmers, deflector, and the knife edge are mounted on motorized
translation stages in the $x$--$y$ plane. The individual positions can therefore be adjusted
perpendicular to the molecular beam propagation direction using linear feedthroughs and stepper
motors (Thorlabs, DRV014), which can be controlled remotely. The maximum travel range for the whole
molecular beam is $\ordsim11$ and $\ordsim12$~mm in $x$- and $y$ direction, respectively. The
implemented motors enable the alignment and vertical scanning of the molecular beam with respect to
any general photon source, enabling the investigation of different sections of the
quantum-state-dispersed molecular beam.

An Even-Lavie valve~\cite{Even:EPJTI2:17} is used to deliver the molecular beam by a supersonic
expansion of 10--100~bar helium with typically a few mbar of sample molecules into
vacuum~\cite{Trippel:MP111:1738}. The valve temperature can be adjusted between room temperature and
\celsius{250} to control the vapor pressure of the sample under investigation and can be operated
with a repetition rate of up to 500~Hz. The valve is mounted on a semi-motorized DN63~CF 3D
manipulator (Caburn-MDC E-PSM-2504) with a travel range of 7~cm along the molecular beam propagation
direction. The first conical skimmer (Beam Dynamics, model 50.8, $\varnothing=3$~mm) is used for
skimming the molecular beam and to obtain differential pumping. The distance between the first
skimmer and the valve can be adjusted between 9 and 16~cm by moving the valve along the molecular
beam propagation direction. The first skimmer is mounted on a home-built flange with a built-in 2D
translation stage to adjust the skimmer position in the $x$- and $y$ directions. A second skimmer
(Beam Dynamics, model 40.5, $\varnothing=1.5$~mm) is placed 13.01~cm behind the first skimmer tip
for further collimation of the molecular beam. An electrostatic $b$-type
deflector~\cite{Kienitz:JCP147:024304}, which disperses the molecules in the molecular beam with
respect to their quantum states~\cite{Chang:IRPC34:557, Filsinger:JCP131:064309,
   Filsinger:PRL100:133003, Trippel:PRA86:033202, Teschmit:ACIE57:13775} is located 4.4~cm behind
the tip of the second skimmer. Both electrodes of the electrostatic deflector are connected through
high-voltage feedthroughs (Pfeiffer 420XST040-30-30-1) allowing voltages of up to $\pm30$~kV on each
electrode. The second skimmer and the electrostatic deflector are mounted together on an $x$--$y$
translation stage to adjust their positions. The electrostatically dispersed molecular beam is then
cut by a vertically adjustable knife edge placed 4.1~cm downstream of the deflector exit. This
allows for both, an improved sample separation and a higher column density of the molecular
beam~\cite{Trippel:RSI89:096110}. The molecular beam is further skimmed by a third conical skimmer
(Beam Dynamics model 50.8, $\varnothing=1.5$~mm) placed 2.37~cm downstream of the knife edge,
providing differential pumping against the VMI chamber. This third skimmer is again mounted on a 2D
translation stage. The molecular beam enters the VMI chamber after the third skimmer. The distance
between the tip of the third skimmer and the interaction region is 27.8~cm. The state-selected
molecules are ionized by the photons used for the investigation inside the double-sided VMI
spectrometer~\cite{Eppink:RSI68:3477}. The resulting electrons and ions are velocity-mapped onto a
120~mm (Photonis Model: APD 3 PS 120/32/25 19 I 60:1 NR 10''FM P47) and a 75~mm (Photonis Model: APD
3 PS 75/32/25/8 I 60:1 NR MGO 8''FM P47) diameter position-sensitive detectors, respectively. Each
detector consists of three z-stacked multi-channel plates (MCPs) and a fast phosphor screen with a
decay time on the order of 100~ns (P-47)~\cite{Winter:RSI85:0034}. Both detectors are captured by
time and position-sensitive Timepix3 cameras built by Amsterdam Scientific Instruments based on
homebuilt chips~\cite{Bromberger:JPB55:144001,Heijden:JINST12:C02040}. The molecular beam enters the
beam dump chamber through a $\varnothing=2$~cm aperture after passing the VMI. The beam dump chamber
pump is mounted with a 22 mm offset regarding the z-axis, allowing the molecular beam to directly
hit onto the rotors to prevent scattering of the direct beam back into the VMI chamber.

\subsection{Double-sided VMI}
\label{sec:Double-sided VMI}
The double-sided VMI setup specifically designed for eCOMO is illustrated in
\autoref[a]{fig:VMI-setup} along with its relevant dimensions and the electrode labels.
\begin{figure*}
   \includegraphics[width=\linewidth]{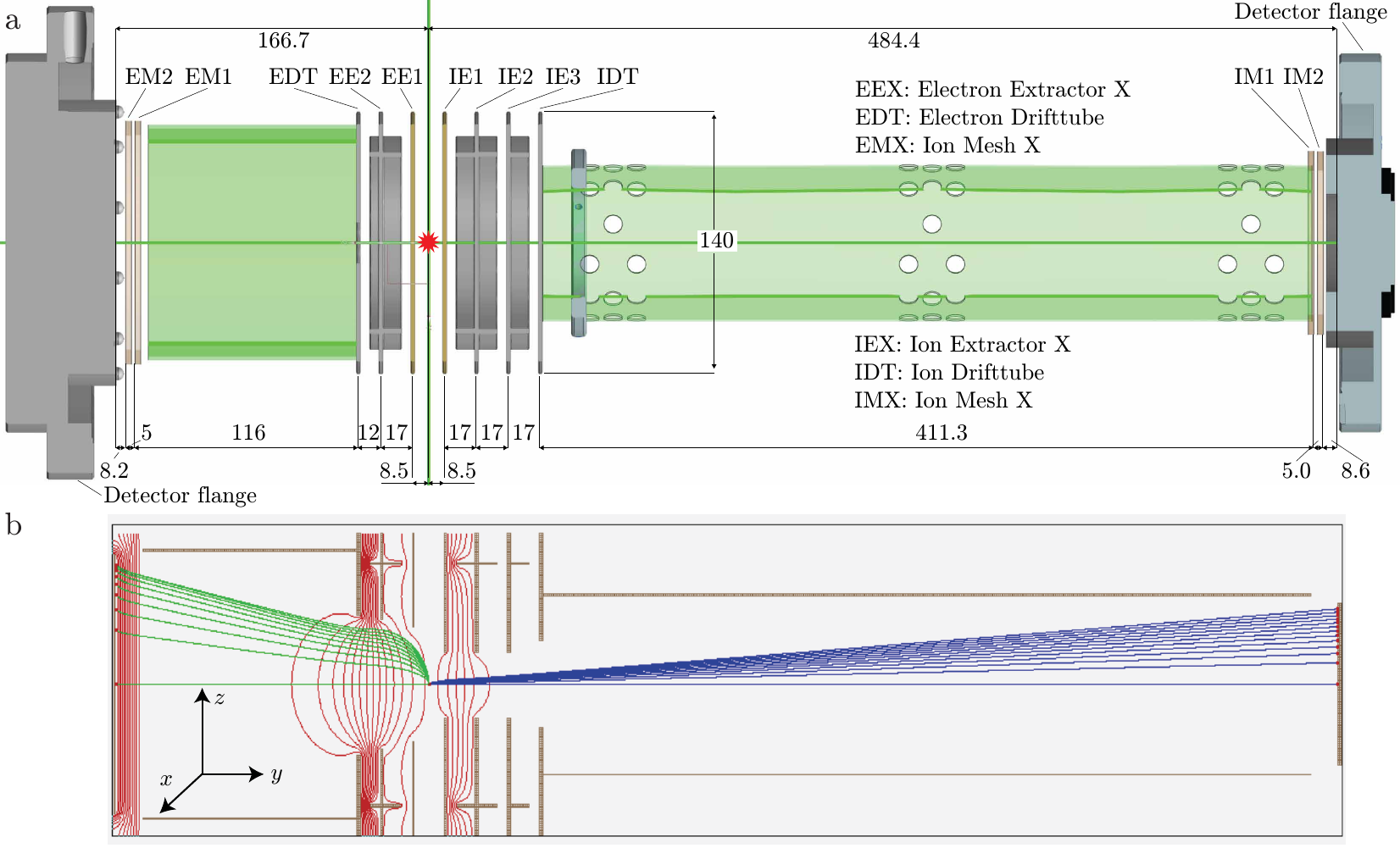}%
   \caption{(a) Schematic of the double-sided-VMI setup cut along the $y$-$z$ plane, containing the
      symmetry axis. The left side represents the electron detector, while the right side represents
      the ion detector. All distances are in mm. Typical voltage settings for each electrode can be
      found in \autoref{tab:voltage_setting}. The interaction region is indicated by the red star.
      (b) Simulation for the large velocity range mode calculated with SIMION. Equipotential lines
      are displayed in red. The trajectories of electrons are depicted in green, while ion
      trajectories are represented in blue. Electron kinetic energies parallel to the detector
      surface were chosen in steps of 70~eV from 0 to 630~eV and ion kinetic energies parallel to
      the detector surface in steps of 1~eV from 0 to 12~eV.}
   \label{fig:VMI-setup}
\end{figure*}
It allows for the simultaneous detection of electrons and ions and evolved from the initial concept
of velocity imaging in photoionization coincidence studies~\cite{Takahashi:RSI71:1337}. The VMI's
adaptable design permits both, vertical and horizontal installations within the VMI chamber,
facilitating flexible use in different laboratories and photon-science facilities. By default, the
eCOMO spectrometer is installed vertically in the VMI chamber. Alternatively, the spectrometer can
be installed horizontally along the molecular beam direction replacing the beam dump. In this case,
only the long side of the spectrometer can be used for either electrons or ions. Furthermore, this
arrangement results in a higher background gas pressure inside the VMI, because the atoms and
molecules only leave this region through the additional holes in the drift tubes after colliding
with the detector and the walls inside the spectrometer. The total conductance of the relevant part
of the VMI for helium is $\ordsim2500$~l/s and the corresponding total relevant volume is
$\ordsim3.6$~l. This results in a $1/e$ time constant on the order of 1~ms. For the same
molecular-beam-valve repetition rate as used here the resulting background-pressure increase is
$\ordsim 6\times10^{-10}$~mbar for helium.

The VMI's point of origin is defined in the middle between the electrodes EE1 and IE1 as indicated
by the red star. This is the location where the molecular beam interacts with the
photons. Electrodes EE1 and IE1 are fabricated from oxygen-free copper (OF-Cu) with at least 5~\um
of gold plating to obtain a higher work function compared to stainless steel for experiments using
high energetic photons, \eg, UV light. The remaining electrodes are made of stainless steel (1.4429
ESU), with surfaces finely polished. Additionally, a rim structure is designed around the outer edge
of each electrode to reduce the influence of the VMI holder on the shape of the electric field in
the center of the spectrometer. All electrodes have an outer diameter of 140~mm and are separated by
ceramic spacers (Macor) as electrical insulators. A cylindrical shield surrounding the VMI, depicted
in green, with holes for the photons, molecular beam, potential diagnostic paddles for x-ray
experiments as well as for pumping, is composed of $\mathrm{\mu}$-metal to provide shielding from
stray fields and external magnetic fields. Double-layered meshes EM1--EM2 and IM1--IM2, with 80\%
transmission for each layer, are positioned on both sides of the spectrometer to ensure homogeneous
drift tube fields close to the detector surfaces. The first mesh on both sides with respect to the
interaction region is conductively connected to the corresponding drift tube. The VMI is completely
mounted on the DN200~CF ion detector flange. Its design allows for straightforward installation or
replacement entirely from the ion detector side. The electrode connections extend through wires to
the DN200~CF flange, where they proceed to the exterior via electrical feedthroughs
(CF16-SHV20-SH-SE-CE-NI, Vacom) with a maximum rated voltage of 20~kV each.

The spectrometer can operate in diverse photo-electron and photo-ion detection modes by adjusting
the voltages across the electrodes as detailed in \autoref{tab:voltage_setting}.
\begin{table}
    \centering\footnotesize%
    \begin{tabular}{lccc}
        \hline
        Imaging mode & SMI & SVR--VMI & LVR--VMI\\
        \hline\hline
        Electron mesh 2 (EM2)& 0 & -134 & 6864\\
        \hline
        Electron drift tube (EDT)& 0 & 1956 & 9816\\
        \hline
        Electron extractor 2 (EE2)& 0 & 490 & 673\\
        \hline
        Electron extractor 1 (EE1)& 0 & -277 & -553\\
        \hline
        Ion extractor 1 (IE1)& 0 & -370 & -593\\
        \hline
        Ion extractor 2 (IE2)& -420 & -821 & -4672\\
        \hline
        Ion extractor 3 (IE3)& -1600 & -2373 & -4717\\
        \hline
        Ion drift tube (IDT)& -1600 & -2380 & -4800\\
        \hline
        Ion mesh 2 (IM2)& -1600 & -821 & -4672\\
        \hline
    \end{tabular}
    \caption{The voltage settings (in V) required for the corresponding electrodes, as displayed in
       \autoref[(a)]{fig:VMI-setup}, to enable the operation of the double-sided VMI in the
       specified modes. See text for details.}
    \label{tab:voltage_setting}
\end{table}
This table lists, as examples, the implementation of a spatial map imaging (SMI)
mode~\cite{Stei:JCP138:214201} -- optimized exclusively for ions -- and two VMI modes for different
maximum energies for the velocity components parallel to the detection plane $E_\text{p}$. The VMI
modes images both, electron and ion velocities. Each voltage setting is tailored for specific
applications. In the SMI mode, the projected position on the detector correlates with the ion's
initial position and is in the first order unaffected by its initial velocity, enabling precise
imaging of the ionizing region. This functionality is particularly advantageous for the alignment of
the experimental setup, ensuring accurate positioning of the light focus and the molecular beam.
Conversely, the VMI mode allows electrons or ions with identical velocity vectors to converge to a
single spot on the corresponding detector, in the first order irrespective of their initial
positions. Different VMI modes allow different maximum kinetic energies to be collected on the
detector, which can be adjusted depending on the experimental situation. The detection range for the
electron kinetic energy parallel to the detector surface extends up to 700~eV in the
large-velocity-range VMI (LVR--VMI) mode. Single-charged ions can be mapped here with a parallel
energy up to 12~eV. The range for the parallel electron kinetic energy extends up to 100~eV and for
single-charged ions up to 10~eV in the small-velocity-range VMI mode (SVR--VMI).

The combination of MCP and phosphor screen amplifies each individual ion and electron after the
particles hit the detector, by creating an avalanche of electrons that then scintillates on the
phosphor screen. The light-flashes are captured with a Timepix3 camera~\cite{Zhao:RSI88:113104,
  Bromberger:JPB55:144001} operated and controlled by our open source library
PymePix~\cite{AlRefaie:JINST14:P10003, pymepix:gitlab}. PymePix is also utilized to extract the raw
physics events from the Timepix3 data stream, converting them into centroided events recognized as
single particle events~\cite{Bromberger:JPB55:144001}. The data-driven cameras, with a temporal
resolution of $\ordsim 1.6$~ns, enable us to operate in multi-mass detection mode and to directly
obtain the VMI images of all fragments by slicing the TOF coordinate in software during
analysis. The performance evaluation of the Timepix3 camera is beyond the scope of this article and
was extensively discussed in other publications~\cite{Bromberger:JPB55:144001, Cheng:RSI93:013003,
  Bromberger:jinst19:P11008}. The relative positions and relative TOFs measured by the cameras can
be converted to velocities in VMI conditions.

The energy resolution of the double sided VMI was determined through SIMION simulations. It is in
the first approximation linearly dependent on the source size so the volume in which the ions and
electrons are produced. In addition, it depends on the specific mode of spectrometer operation. In
our specific simulation we used a cylinder for the focal volume with the length given by the
diameter of the molecular beam $D=2$~mm, and the radius given by the standard deviation of the
transverse laser intensity distribution $\sigma_{r}=21$~\um resembling our experimental conditions.
Taking LVR--VMI as an example, we obtain $\Delta E_\text{p}/E_\text{p} = 0.004$ for the ion side, which
is to first order constant for all energies that can be mapped. This results in an energy resolution
of $\Delta E_\text{p}=0.02$~eV for, \eg, a parallel ion energy of $E_\text{p}=5$~eV. For electrons,
using the same settings, we obtain $\Delta E_\text{p}/E_\text{p} = 0.003$. Both resolutions are in
good agreement with results obtained on similar systems~\cite{Rading:AS8:998}. Further settings to
meet special experimental requirements are possible, but must be again found by simulations.

\subsection{Portability and Installation}
\label{sec: Portability and Installation}
The eCOMO setup is designed for easy and convenient transport between different light sources, which
is facilitated by four integrated wheels. Once eCOMO is positioned, four adjustable home-built
support feet connected to a screw jack with standing screw (Nozag NSE25), capable of both,
horizontal and vertical adjustments, allow precise 2D positioning with a travel range of
$\pm3$~cm. The overall height of the interaction center can be adjusted between 103 and 135~cm. The
source, deflector, and VMI chambers are mounted on profile guide rails attached to the stainless
steel platform by guide carriages, allowing further adjustments along the molecular beam
direction. The eCOMO setup incorporates all necessary equipment, as depicted schematically in the
upper left corner of \autoref{fig:experimental_setup} in the green highlighted area, except for the
computer used to control eCOMO and acquire data~\cite{AlRefaie:JINST14:P10003}. Integrated within
the setup are all required power supplies with ports allowing quick installation, including three
power cables capable of handling the necessary electrical load, a pair of inlet and outlet pipes for
water circulation to cool the molecular pumps, an exhaust pipe for the vacuum system, and multiple
gas sources such as nitrogen, compressed air, and the molecular carrier gas. The total weight of
eCOMO including the pumps and electronics is $\ordsim1800$~kg. The overall dimensions of eCOMO are
90, 175, and 240~cm in width, height, and length, respectively.

\section{Characterization of the setup}
\label{sec:Characterization of the setup}
\subsection{Molecular beam spatial profiles}
\label{sec:Molecular beam spatial profiles}
\begin{figure}
   \includegraphics[width=\linewidth]{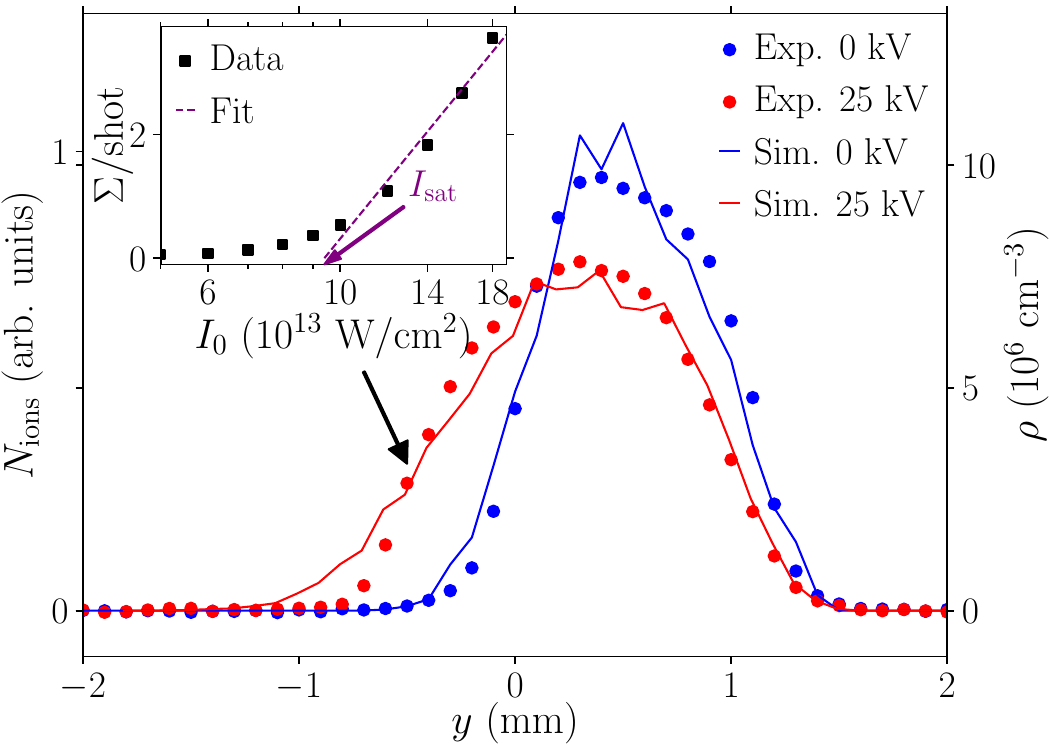}%
   \caption{Measured molecular beam density profiles (points) and simulation (solid lines) for the
      deflector switched off (blue) and at a total voltage of 25~kV (red) applied across the
      electrodes. The inset shows the number of \ch{OCS^{+}} ions $\Sigma$ per shot (black squares)
      as a function of the laser peak-intensity natural logarithm $\ln I_{0}$. The purple dashed
      line represents the fit of the asymptotic slope. The data in the inset was measured at
      $y=-0.5$~mm in the deflected molecular beam profile, marked by the black arrow.}
   \label{fig:molecular_beam_profile}
\end{figure}

We used OCS as a sample to characterize the molecular beam and VMI performance to commission
eCOMO~\cite{Wiese:NJP21:083011, Trippel:PRL114:103003}. OCS was ionized by pulses from a commercial
Ti:sapphire femtosecond laser system (Coherent Astrella) operated at a repetition rate of 1~kHz. The
laser beam was compressed to approximately 40~fs with a standard grating-based compression setup.
The laser beam was directed through a 75~cm focal-length lens, which resulted in a peak intensity of
$I_0\approx3\times10^{14}$~\Wpcmcm. The laser focus was spatially and temporally overlapped with the
molecular beam in the interaction center of the VMI spectrometer. The pulsed molecular beam was
produced by expanding a mixture of 100~bar helium and $\ordsim10$~mbar OCS. The gas mixture was
expanded into the source chamber through the Even–Lavie valve operated at a temperature of
\celsius{50} and at a repetition rate of 250~Hz. Laser shots without the molecular beam were used
for background measurements. Optimized operation conditions of the valve allowed for the generation
of a dense ($>10^{7}~\invccm$) molecular beam in the interaction volume. The OCS sample was
displaced toward negative $y$ positions when a total voltage of 25~kV was applied across the
deflector, \ie, \emph{de facto} the populated quantum states of OCS exhibited a strong-field-seeking
character at the relevant electric field strengths in the deflector. The measured deflection profile
of OCS is shown in \autoref{fig:molecular_beam_profile} as red dots.
Solid lines represent simulated OCS beam profiles, computed through Monte-Carlo sampling combined
with trajectory calculations that incorporate the geometrical constraints of the mechanical
apertures in the experimental setup. These simulated deflection profiles of OCS match the
experimental data, assuming an initial rotational temperature of $T_\text{rot}=0.5$~K for the
molecular beam entering the deflector. The results affirm that the eCOMO setup can generate cold and
stable molecular beams.

The OCS beam sample density was determined using a strong-field ionization
model~\cite{Hankin:PRA64:013405, Wiese:NJP21:083011}. The asymptotic slope of an integral ionization
signal $\Sigma(I_{0})$, with respect to natural logarithm of the peak intensity $\ln I_{0}$
approaches a limiting value as $I_{0}$ approaches infinity. This slope can be expressed as:
\begin{equation}
\lim_{ I_{0} \to \infty} \left(\frac{d\Sigma }{d\ln I_{0} } \right) = 2\pi \alpha \sigma _{r}^{2} D\rho
\end{equation}
where $\rho$ represents the sample density, $\sigma _{r}$ the standard deviation of the transverse
laser intensity distribution, $\alpha$ the instrument sensitivity, $D$ the length of the focal
volume in the molecular beam, and $\Sigma$ the total number of detected ions. The inset of
\autoref{fig:molecular_beam_profile} illustrates the parent-ion signal count $\Sigma$ per shot as a
function of the laser peak intensity $I_0$ on a semi-logarithmic scale. The data was measured at
$y=-0.5$~mm in the deflection profile as indicated by the black arrow. The asymptotic slope and
saturation onset were obtained by fitting a straight line through the highest peak-intensity points,
resulting in a saturation onset of $I_\text{sat} = (9.4 \pm 0.3_\text{stat} \pm 1.9_\text{syst})
\times 10^{13}~$\Wpcmcm. The subscripts indicate statistical and estimated systematic errors. This
is in good agreement with the previously reported saturation intensity of OCS measured under similar
experimental conditions given by $I_\text{sat} = 7.2 \times
10^{13}~$\Wpcmcm~\cite{Wiese:NJP21:083011}. To extract the sample density we determined the combined
sensitivity of the meshes and the MCP detector to be $\alpha = 35.9\% \pm 1.3\%$ as described in
\autoref{sec:VMI}. In addition we use $\sigma _{r}=21$~\um and $D=2$~mm for the laser intensity
standard deviation and the molecular beam width, respectively. The resulting sample density at
$y=-0.5$~mm is given by $\rho = (2.86 \pm 0.58_\text{stat} \pm 0.57_\text{syst}) \times
10^{6}~\invccm$. The maximum density is $\rho \approx 10^{7}~\invccm$ in the direct beam as shown in
\autoref{fig:molecular_beam_profile}, which is one order of magnitude lower than previously reported
on a similar experimental setup~\cite{Wiese:NJP21:083011}, which is attributed to a lower
concentration of OCS in the helium mixture.

\subsection{Time-of-flight mass spectrum (TOF-MS)}
\label{sec:Time-of-flight mass spectrum(TOF-MS)}
\begin{figure}
   \includegraphics[width=\linewidth]{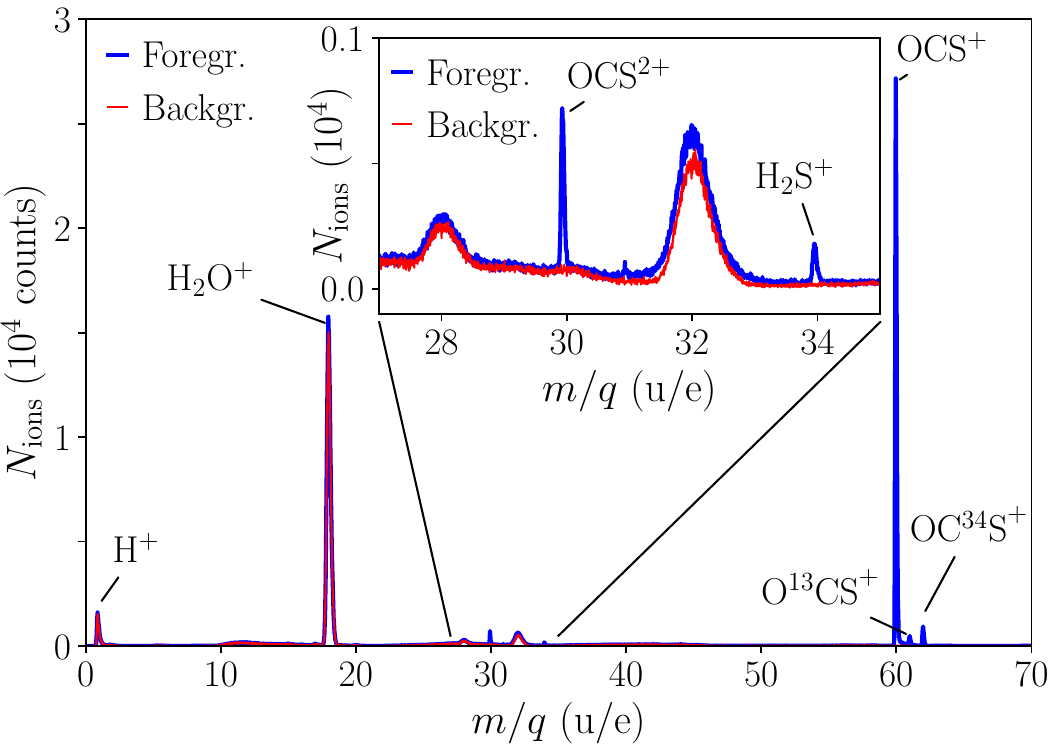}%
   \caption{TOF-MS spectrum of ions originating from the molecular beam containing OCS (blue) and
      from VMI-chamber background gases (red). The inset provides a magnified view of the specific
      region where OCS fragment ions and more-highly-charged ions occur.}
   \label{fig:TOF}
\end{figure}

\autoref{fig:TOF} shows the time-of-flight mass spectrum (TOF-MS) resulting from strong-field
ionization of OCS and from the background gas in the VMI chamber measured with the Timepix3
camera~\cite{Zhao:RSI88:113104, AlRefaie:JINST14:P10003, Bromberger:JPB55:144001}.
The figure displays two distinct regions: The first region depicts the \ch{OCS} parent ion and its
isotopologues O$^{13}$CS and OC$^{34}$S, which correspond to \mq{60}, \mq{61}, and \mq{62},
respectively. The respective signal strength is in good agreement with the corresponding natural
abundance of OCS's most abundant atmospheric isotopologues~\cite{Yousefi:JQSRT238:106554}. The
isotopologues with masses \mq{63}, and \mq{64} are also present in the data, but are not visible in
the graph shown because of their natural abundance below 0.1\%. The second feature is the presence
of fragments, as also seen in the zoomed-in subfigure of the graph. Two broad peaks correspond to
\ch{CO^+} with \mq{28} and \ch{S^+} with {\mq{32}} ions, which exhibit a broad ion velocity
distribution due to Coulomb explosion of \ch{OCS^{2+}}. The broad velocity distribution is directly
reflected in the broad TOF distribution. The relatively strong background here originates from
impurities of \ch{O_2} and \ch{N_2} gases present in the VMI chamber and the molecular
beam. Additionally, two narrow peaks are observed in the inset, which correspond to doubly ionized
\ch{OCS} and singly ionized \ch{H_2S}, which is an impurity also present in the molecular beam. The
significant water ion peak can also be mostly attributed to poor background pressure conditions,
which at the time, were not fully optimized and given by $P_\text{VMI} = 4 \times 10^{-9}$~mbar when
the molecular beam was switched off. Additionally, there are small contributions of water ions that
arose from impurities in the molecular beam. Recent modifications of the setup showed considerable
improvement, achieving pressures as low as $P_\text{VMI} = 3 \times 10^{-11}$~mbar. Furthermore, the
purity of the gas line was also improved by installing a cooling thermostat (ECO RE 1050 S, LAUDA).

\subsection{Performance of the double-sided VMI}
\label{sec:VMI}

In order to experimentally characterize the double-sided VMI, both electrons and ions from the
molecular beam containing OCS were examined after strong-field ionization. \autoref[a]{fig:VMI} and
\autoref[c]{fig:VMI} show the background subtracted cation-fragment velocity maps for \mq{28} and
\mq{32}, respectively. The laser-peak-intensity here was $I_0 = 2.5 \times 10^{14}$~\Wpcmcm.
\begin{figure}
   \includegraphics[width=\linewidth]{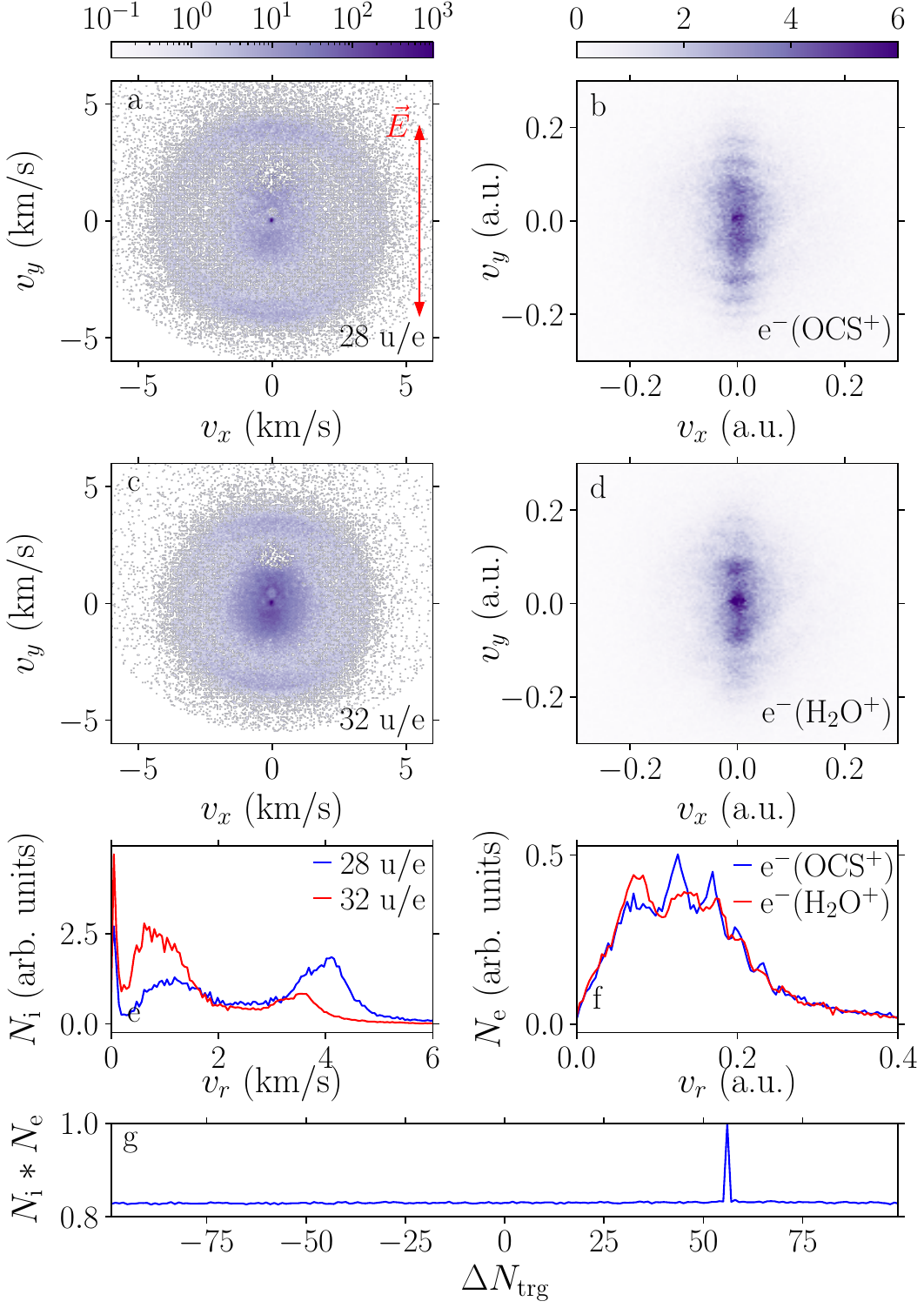}
   \caption{(a) and (c) Projected ion velocity maps for \mq{28} and for \mq{32} fragments,
     respectively. The laser polarization is indicated by the red arrow. (b) and (d) Projected
     electron velocity maps for electrons correlated with \ce{OCS^+} and \ce{H_{2}O^+},
     respectively. (e) and (f) Projected radial velocity distributions for ions and electrons,
     respectively. (g) Ion-electron Timepix3-camera-trigger correlation. See text for more details.}
   \label{fig:VMI}
\end{figure}
The velocity maps show the commonly known features of \ce{OCS} fragments after strong field
ionization~\cite{Karamatskos:FD228:413, Trippel:PRL114:103003}. The central circular parts of both
velocity maps are attributed to fragmentation to \ce{CO^+} and \ce{S^+} following single ionization.
The intense central spots are attributed to N$_2^+$ and O$_2^+$ in the \mq{28} and \mq{32} velocity
maps, respectively. These were impurities present in the molecular beam. The outer rings are
ascribed to the Coulomb explosion channels \ce{OCS + $nh\nu$ \rightarrow OCS^{+2} \rightarrow CO^+ +
   S^+}. The reduced counts right above the central regions are attributed to strong signals from
background ions such as \ce{H_2O^+} arriving shortly before the selected ions in combination with
the dead time of the Timepix3 camera, which was on the order of 1~\us in our
case~\cite{Bromberger:JPB55:144001}. In addition, the detector could be less sensitive there because
all background ions hit this area. The projected radial velocity distributions are shown in
\autoref[e]{fig:VMI}. The multiple channels corresponding to the VMIs are visible again.

The ion information can be used to determine the electron velocity map corresponding to specific
ionization or fragmentation channels. To do so, both Timepix3 cameras have to be first synchronized
as they independently record electrons and ions. To correlate the ion and electron cameras we
simultaneously took ion and electron data at low laser intensities ($I_0 = 3 \times
10^{13}$~\Wpcmcm) at an average event rate of 0.02~ions or electrons per shot. The number of ions
$N_\text{i}(N_\text{trg;i})$ or electrons $N_\text{e}(N_\text{trg;e})$ were assigned to the
camera-specific trigger numbers $N_\text{trg;i}$ and $N_\text{trg;e}$, respectively. The normalized
convolution $N_\text{i} \ast N_\text{e}$ of this signal is shown in \autoref[g]{fig:VMI} as a
function of the relative trigger number $\Delta N_\text{trg} = N_\text{trg;i}-N_\text{trg;e}$. The
peak indicates the required relative trigger number shift for both cameras to be synchronized.

To demonstrate ion-electron correlations, we ionized OCS with a laser-peak-intensity of
$I_0 = 8 \times 10^{13}$~\Wpcmcm to have both, parent ions and fragments present. In this case, the
mean number of detected ions per shot was $\ordsim2$ so specific ions and electrons from single
laser shots can not be correlated directly. We can, however, obtain the statistically averaged
electron image using the covariance of the electron velocity maps with the boolean operation on a
specific ion detected~\cite{Allum:CommChem5:42}. \autoref[b]{fig:VMI} displays the photoelectron
velocity distribution after the covariance analysis where an \ce{OCS^{+}} was measured in
coincidence. The figure reveals intricate angular and radial structures as previously
reported~\cite{Wiese:NJP21:083011}. Specifically, the momentum distributions exhibit wide radial
patterns that repeat at integer multiples of the photon energy, which is commonly referred to as
above-threshold ionization (ATI). A detailed discussion about the structure of the OCS electron
momentum map is beyond the scope of this manuscript and further information can be found
elsewhere~\cite{Wiese:NJP21:083011}. \autoref[d]{fig:VMI} displays the photoelectron momentum
distribution accordingly when an \ch{H_2O^{+}} from ionization of the background gas was measured in
coincidence with an electron. A clear difference in the image structure is observed in comparison
with \autoref[b]{fig:VMI}, which is attributed to the different ionization channels. The differences
become even more obvious in the electron's speed distribution projected onto the detector shown in
\autoref[f]{fig:VMI}. In comparison with previous work~\cite{Wiese:NJP21:083011}, our short data
acquisition time led to smaller experimental statistics, resulting in less detailed ATI structures.
Further discussion about the underlying reason with respect to the differences in the electron
velocity distributions is again beyond the scope of the manuscript. The clear differences between
the two electron velocity maps however clearly show that eCOMO allows to correlate electron and ion
momenta.

The low-intensity correlation measurements used for \autoref[g]{fig:VMI} enabled us furthermore to
obtain the detection efficiency for electrons and ions in our setup (see SI for details). From our
data we obtain a final total detection efficiency of $\alpha_\text{i}=35.9\% \pm 1.3\%$ and
$\alpha_\text{e}=35.4\% \pm 1.3\%$. Assuming a detection efficiency of the MCP given by
$\alpha_\text{MCP}=50$\% we obtain a transparency for a single mesh $\alpha_\text{mesh}=83.7\% \pm
1.5\%$, with the estimated transmission of 80\%. The deviation is attributed to the fact that the
effective transmission of the second grid is higher due to the shadow cast by the first grid.

\section{Conclusions}
\label{sec:conclusions}

In summary, eCOMO was developed, set up, and commissioned using experiments with an ultrashort-pulse
table-top laser system. The performance of the cold ($T_\text{rot}\le1$~K), pulsed, and dense
($\rho \approx 1 \times 10^{7}~\invccm$) molecular beam was characterized. SIMION simulations of the
double-sided VMI indicated that under experimental conditions, an energy resolution of
$\Delta E/E < 0.4 \%$ is achievable. We showed that the combination of the double-sided VMI and two
synchronized Timepix3 cameras at eCOMO works and that the individual ions can be separated in the
data analysis based on their mass over charge ratio. Furthermore, specific molecular dynamics can be
separated for systems with many fragmentation channels as shown here for OCS after ionization. Using
a covariance analysis, electron VMIs and corresponding projected electron velocity distributions
were assigned to specific mass over charge ratios. The absolute detection efficiency of the setup
was measured by coincidence methods to $\alpha_\text{i}=35.9\% \pm 1.3\%$ and
$\alpha_\text{e}=35.4\% \pm 1.3\%$ for ions and electrons, respectively. The use of the $b$-type
electrostatic deflector allows to separate specific rotational quantum states, conformers and
molecular cluster species with different effective-dipole-moment-to-mass
ratios~\cite{Chang:IRPC34:557, Kienitz:JCP147:024304}. This feature of the eCOMO setup has great
potential in various areas, such as the separation and purification of complex clusters for chemical
dynamics studies~\cite{Onvlee:NatComm13:7462}.

The eCOMO setup exhibits versatile applications in the production of purified cold gas-phase
molecular beams and high-precision ion-electron coincidence and covariance measurements. It can be
readily adapted to a wide range of photon-source facilities, making it suitable for diverse research
activities spanning atomic, molecular, and cluster physics, as well as materials science, energy
science, chemistry, and biology.

\section{Acknowledgments}
We thank Ilya Maksimov for his experimental contributions in an early phase of the project and
Daniel Gusa and the electronics workshop at DESY for technical support.

We acknowledge financial support by Deutsches Elektronen-Synchrotron DESY, a member of the Helmholtz
Association (HGF), also for the provision of experimental facilities and for the use of the Maxwell
computational resources operated at DESY. This research was supported by the Center for Molecular
Water Science (CMWS) in an Early Science Project. The research was further supported by the Cluster
of Excellence ``Advanced Imaging of Matter'' (AIM, EXC~2056, ID~390715994) of the Deutsche
Forschungsgemeinschaft (DFG), the Helmholtz Foundation through funds from the Helmholtz-Lund
International Graduate School (HELIOS, HIRS-0018), the German Federal Ministry of Education and
Research (BMBF) and the Swedish Research Council (VR~2021-05992) through the Röntgen-Ångström
cluster project ``Ultrafast dynamics in intermolecular energy transfer: elementary processes in
aerosols and liquid chemistry'' (UDIET, 05K22GUA), and the European Union's Horizon 2020 research
and innovation program under the Marie Skłodowska-Curie Grant Agreement ``Molecular Electron
Dynamics investigated by Intense Fields and Attosecond Pulses'' (MEDEA, 641789). We acknowledge
CAMP\textbf{@}FLASH at DESY for providing some material for the spectrometer.

\section{Data availability}
The data recorded in the laser lab at DESY are available from the corresponding author upon
reasonable request.

\section{Code availability}
The code for the data analysis is available from the corresponding author upon reasonable request.

\bibliography{string,cmi}

\onecolumngrid
\listofnotes  
\end{document}


\title{\emph{Supporting Information:} \\ A versatile and transportable endstation for controlled molecule experiments}
\author{Wuwei~Jin}\cfeldesy\uhhphys%
\author{Hubertus~Bromberger}\cfeldesy%
\author{Lanhai~He}\jlu\cfeldesy%
\author{Melby~Johny}\cfeldesy\uhhphys\uhhcui%
\author{Ivo~S.~Vinkl\'{a}rek}\cfeldesy
\author{Karol~Długołęcki}\cfeldesy
\author{Andrey~Samartsev}\cfeldesy
\author{Francesca Calegari}\cfeldesy\uhhphys\uhhcui
\author{Sebastian~Trippel}\stemail\cfeldesy\uhhcui%
\author{Jochen~Küpper}\jkemail\cmiweb\cfeldesy\uhhphys\uhhcui
\maketitle%


\section{Vacuum system}
\label{SI:sec:vacuum system}
\begin{figure*}
   \includegraphics[width=1.0\textwidth]{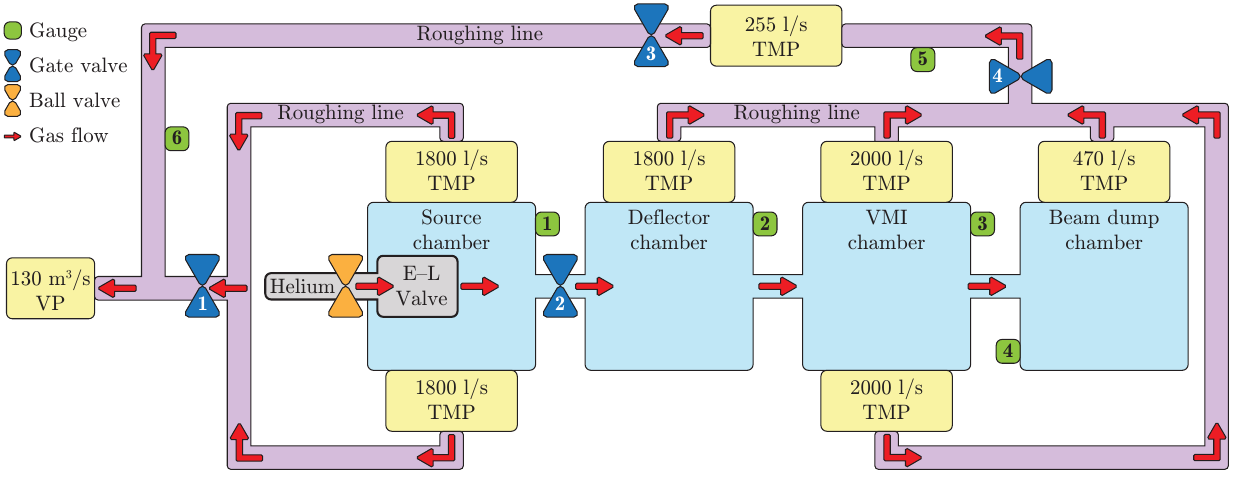}%
   \caption[Vacuum system]{Schematic of the eCOMO vacuum system together with the high pressure gas
     line. Gauges and pumps are indicated as green and yellow rectangles, respectively. Red arrows
     depict the gas flow. Ball and gate valves are indicated by orange and blue symbols,
     respectively. The high-vacuum part of the setup is depicted in light blue. The roughing lines
     are shown in light purple. The high-pressure line and the Even-Lavie valve (EL valve) are
     shown by the gray shaded structure.}
   \label{fig:vacuum system}
\end{figure*}

\autoref{fig:vacuum system} illustrates the vacuum setup, including the high-pressure gas line of
the eCOMO system.
The source-, deflector-, VMI- and beam-dump chambers are indicated as the light blue structure. The
gas flow is depicted by red arrows starting from the Even-Lavie (EL) valve in the source chamber.
The turbomolecular pumps are indicated as yellow rectangles and are discussed in the main
manuscript. The roughing-line pump is indicated as OTB. The automatic gate valves are shown as blue
symbols. Here, gate valves 1 and 4 are DN40-ISO-KF mini gate valves (VAT 01232-KA44). Gate valve 3
is a DN16~ISO-KF mini gate valve (VAT 01224-KA44). Gate valve 4 indicates an updated home-built
automatic DN250~CF gate valve~\cite{Kuepper:RSI77:016106}, which allows separating the source from
the deflector chamber. This gate valve is driven by a solenoid valve (Festo
VUVS-LK20-M52-AD-G18-1C1-S). All gate valves are controlled by a home-built NIM-based controller.
Vacuum monitoring is accomplished by the gauges shown in \autoref{fig:vacuum system} as green
rectangles (Pfeiffer Vacuum PKR 251, PBR260, and IKR270). All gauges are controlled and monitored by
a controller (Pfeiffer TPG 366).

\section{Interlock system}
\label{SI:sec:interlock system}
\begin{figure*}
  \includegraphics[width=1.0\textwidth]{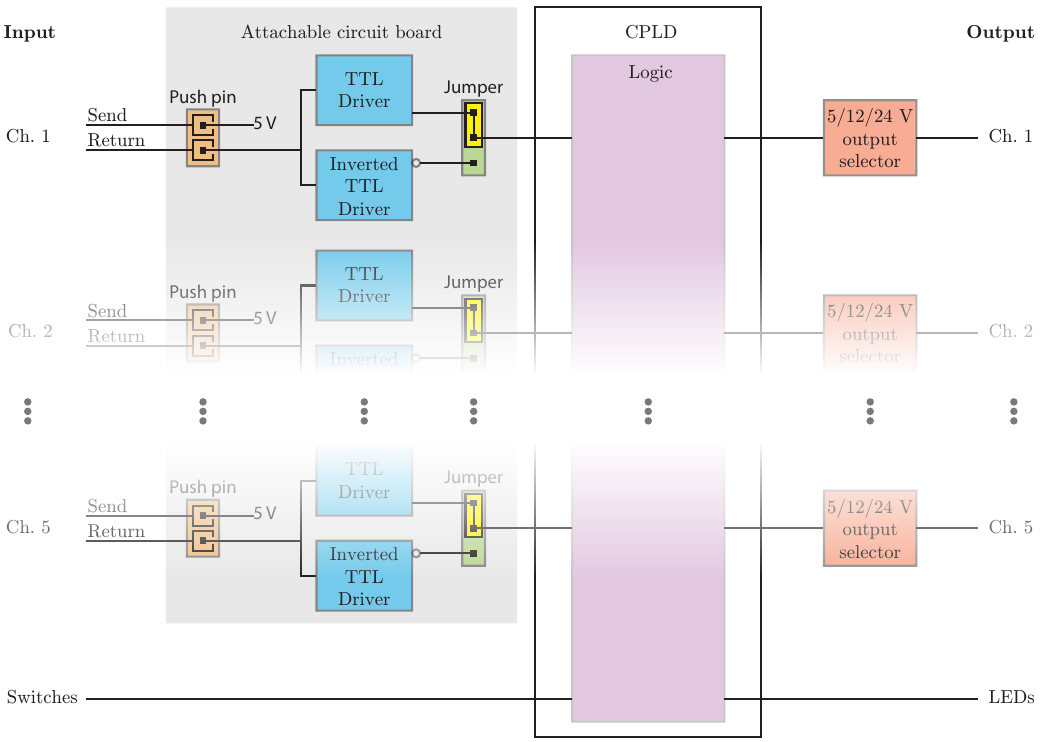}%
  \caption{General concept of the interlock controller. The middle section is hidden for space
    reasons.}
  \label{fig:interlock_concept}
\end{figure*}
The general concept of the interlock controller logic is illustrated in
\autoref{fig:interlock_concept}. The box has 5 two-wire LEMO inputs and 5 one-wire LEMO outputs. One
of the two-wire input wires delivers 5~V (``Send''). The second wire is used to drive the
transistor-transistor logic (TTL) driver integrated circuits (ICs) labeled as ``Return'' in
\autoref{fig:interlock_concept}. All input channels are connected via push pins with an attachable
circuit board in the interlock controller. The push pins can be used to select a specific pin
assignment of the input connectors with respect to Send and Return. The Return signal drives two
ICs, a TTL driver, and an inverted TTL driver, respectively. In this way, the input logic can be
inverted making use of the jumpers right behind the TTL drivers. The TTL driver ICs serve also as a
protection of the interlock controller regarding over voltages. The attachable circuit board can be
easily replaced if necessary since it is connected to a main circuit board via push pins. A complex
programmable logic (CPLD) device is used to logically connect the input signals and interlock
hardware switches with the output channels and LEDs. In this way, the logical program can be changed
quickly. The output voltage can be switched between 5, 12, and 24~V depending on the specific
requirements of the corresponding device attached.

\begin{figure*}
  \includegraphics[width=1.0\textwidth]{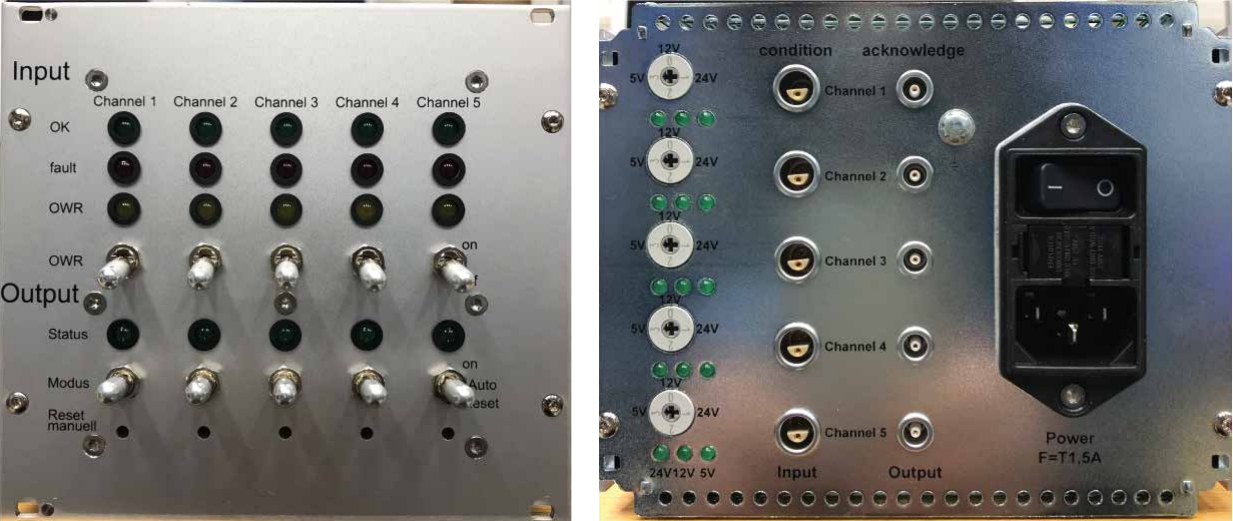}
  \caption{Front- (left) and back-panel (right) of the interlock controller.}
  \label{fig:front-back-panel}
\end{figure*}

The front and back panel of the interlock controller are shown in \autoref{fig:front-back-panel}.
The upper half of the front panel contains LED indicators and switches for the input channels. The
lower half of the front panel contains LED indicators and switches to indicate the status and
control the output channels. Input and output channels are provided via the back plate of the
module. LEDs on the front panel indicate the status of the box depending on the logic of the CPLD.

\begin{figure*}
   \includegraphics[width=1.0\textwidth]{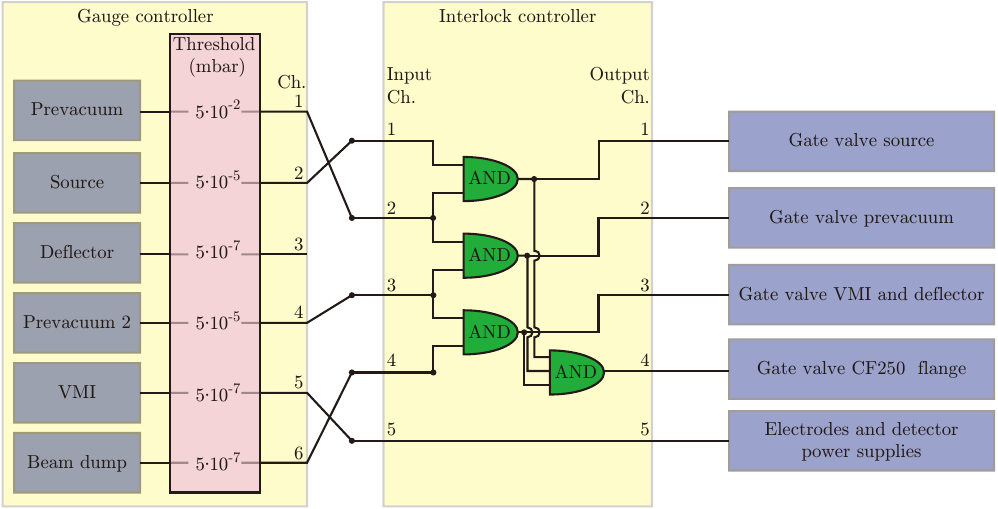}%
   \caption[Interlock system]{Interlock logic used at eCOMO. Thresholds for pressures are set on the
     gauge controller on the left. The corresponding outputs are send to the input of the interlock
     controller. The CPLD of the interlock controller is programmed as illustrated in the
     figure. The logic output of the interlock controller drives the gate valves and the logic for
     the power supplies as indicated on the right.}
   \label{fig:Interlock logic}
\end{figure*}

\autoref{fig:Interlock logic} illustrates how the interlock system monitors and protects the vacuum
system.
The pressure-gauge controller records the vacuum from each gauge and, depending on whether the
respective threshold conditions are met, sends a 5~V trigger signal to the interlock box. The
interlock box then processes these input signals and transmits corresponding output trigger signals
to the gate valves controller box and electrode power supplies to control their operation. In this
special case, a gate valve can only be opened if all vacuum systems connected by this valve are
within the permitted pressure range. Additionally, voltage can only be applied to the electrodes
when the pressure in the VMI chamber is below the threshold.

\section{Detection efficiency}
The low-power correlation measurements used for Figure~5~g in the main manuscript enabled us to
determine the detection efficiency for electrons and ions in our setup. For each detected electron,
we examined whether there was a simultaneous detection of an ion for the same laser shot. The
detection efficiency for ions $\alpha_\mathrm{i}$ and electrons $\alpha_\mathrm{e}$ is given by:
\begin{equation}
   \alpha_\mathrm{i} =\frac{N_\mathrm{i}\cdot N_\mathrm{e} -N_\mathrm{uncor.}}{N_\mathrm{i|e>1}}, \quad
   \alpha_\mathrm{e} =\frac{N_\mathrm{i}\cdot N_\mathrm{e} -N_\mathrm{uncor.}}{N_\mathrm{e|i>1}}
\end{equation}
Here, we have the number of electrons ${N_\mathrm{e}}$, the number of ions ${N_\mathrm{i}}$, the
number of uncorrelated events on both detectors ${N_\mathrm{uncor.}}$, the number of ions with at
least one detected electron ${{N_\mathrm{i|e>1}}}$, and the number of electrons with at least one
detected ion ${{N_\mathrm{e|i>1}}}$. From our data we obtain a final total detection efficiency of
$\alpha_\mathrm{i}=35.90$\% and $\alpha_\mathrm{e}=35.35$\%. Assuming a detection efficiency of the
MCP given by $\alpha_\mathrm{MCP}=50$\% we obtain a transparency for a single mesh
$\alpha_\mathrm{mesh}=83.7$\%, which is in agreement with the nominal transmission of 80\%. The
deviation is attributed to the fact that the effective transmission of the second grid is higher due
to the geometric overlap with the shadow cast by the first grid.

\onecolumngrid

\bibliography{string,cmi}